\documentclass[10pt]{article}
\usepackage{hyperref}
\usepackage{url}
\usepackage{graphicx}
\usepackage{amsmath}
\usepackage{amssymb}
\usepackage[noline]{algorithm2e}
\usepackage{geometry}
\usepackage{longtable}
\usepackage{amsthm}

\usepackage{pgf,tikz}
\usepackage{tkz-euclide}
\usepackage{tkz-graph}

\usetikzlibrary{calc}
\usetkzobj{all}

\usepackage{pgfplots}
\newlength\figureheight 
\newlength\figurewidth 

\newtheorem{lem}{Lemma}

\newtheorem{thm}{Theorem}

\newtheorem{rem}{Remark}

\newtheorem{example}{Example}

\newtheorem*{lem*}{Lemma}
\newtheorem*{pro*}{Proof}
\newtheorem*{thm*}{Theorem}
\newtheorem*{defn*}{Definition}
\newtheorem*{rem*}{Remark}
\newtheorem*{app*}{Appendix}
\newtheorem*{problem*}{Problem}
\newtheorem*{assump*}{Assumption}
\newtheorem*{example*}{Example}
\newtheorem*{cor*}{Corollary}
\newtheorem*{nota*}{Notation}

\usepackage{overpic}

\date{}

\title{Composition Operators, Matrix Representation, and the Finite Section Method: A Theoretical Framework for  Maps between Shapes \footnote{This research was supported by the ERC Starting Grant No. 307047 (COMET).}}

\author{Klaus Glashoff\footnote{Università della Svizzera italiana}, Claus Peter Ortlieb\footnote{University of Hamburg, Germany}}

\begin{document}
\maketitle
%\tableofcontents{ }

\abstract{This paper intends to lay the theoretical foundation for the method of \emph{functional maps}, first presented in 2012 by Ovsjanikov, Ben-Chen, Solomon, Butscher and Guibas  in the field of the theory and numerics of maps between shapes. We show how to analyze this method by looking at it as an application of the theories of \emph{composition operators}, of \emph{matrix representation} of operators on separable Hilbert spaces, and of the theory of the \emph{Finite Section Method}. These are three well known fruitful topics in functional analysis. When applied to the task of modelling of correspondences of shapes in three-dimensional space, these concepts lead directly to \emph{functional maps} and its associated  \emph{functional matrices}.  Mathematically spoken,  functional maps are  composition operators between  two-dimensional manifolds, and functional matrices are infinite matrix representations of such maps. We present an introduction into the notion and theoretical foundation  of the functional analytic framework of the theory of matrix representation, especially of composition operators. We will also discuss two numerical methods for solving equations with such operators, namely, two variants of the \emph{Rectangular Finite Section Method.} While one of these, which is well known, leads to an overdetermined system of linear equations, in the second one the minimum-norm solution of an underdetermined system  has to be computed. We will present the main convergence results related to these methods.
}

\newpage
\section{Introduction}

The theory of \emph{functional maps} and its algorithmic implementations belong to the most fruitful recent ideas in the field of the computation of shape correspondences. The first formulation of this idea appeared in the paper by Ovsjanikov, Ben-Chen, Solomon, Butscher, and Guibas \cite{Ovsjanikov2012} and has since then inspired different research groups in the field of shape correspondence (\cite{ovsjanikov2016computing},\cite{kovnatsky2014functional},\cite{rodola2016partial},\cite{zhang2016functional},\cite{cosmo2016matching}).

In this paper we will look at the basic mathematics underlying this method, and we will show how to build a solid fundament standing on three classical mathematical objects and  related theories:

\begin{description}
\item[1.]Composition operators (\cite{Nordgren1968}, \cite{Nordgren1978}, \cite{singh1993composition},\cite{jafari1998studies},\cite{Takagi1999} )
\item[2.] Matrix representations of operators  (\cite{akhiezer2013theory})
\item[3.] The Finite Section Method (\cite{Boettcher1995},\cite{lindner2006infinite})
\end{description}

We will refer to the relevant literature in these fields, and we will show how the formalism of composition operators etc. which have a rather long history in the realm of Pure Mathematics, may be applied to the relatively young field of maps between shapes.

%%%%%%%%%%%%%%%%%%%%%%%%%%%%%%%%%%%%%%%%%%%%%%%%%%%

\section{Composition operators  }
%%%%%%%%%%%%%%%%%%%%%%%%%%%%%%%%%%%%%%%%%%%%%%%%%%%

One of the first papers on composition operators was the one by Nordgren \cite{Nordgren1978} from 1978. Let us just cite the first two sentences of his paper, as it states in a very lucid way the subject of all subsequent research in this field:

\begin{quotation}
"Let $X$ be a set and suppose $V$ is a vector space of complex valued functions on $X$ under the pointwise operations of addition and scalar multiplication. If $T$ is a mapping of $X$ into $X$ such that the composite $f\circ T$ of $f $ with $T$ is in $V$ whenever $f$ is, then $T$ induces a linear transformation $C_T$ on $V$ that sends $f$ into $f \circ T$."
\end{quotation}

Subsequently, this simple concept of composition operators has been extended and generalized and also specialized to many important function spaces. The case of composition maps between  spaces $L^p(M)$ and $L^q(N)$  with \emph{different} base sets $M$ and $N$ has first been treated in \cite{Takagi1999}.

Given two nonempty sets $M$ and $N$, consider the real vector spaces $\mathcal{F}(M)$ and $\mathcal{F}(N)$ of functions on $M$ and $N$, respectively. For a given  bijective function $$\tau:N\rightarrow M,$$ the \emph{composition operator} 
\begin{eqnarray}\label{def:T1}
T_\tau&:& \mathcal{F}(M) \rightarrow \mathcal{F}(N)
\end{eqnarray}
is defined by 
\begin{eqnarray}\label{def:T2}
T_\tau&=&f\circ \tau,\\
(T_\tau f)(t)&:=&f(\tau(t))
\end{eqnarray}
for every $t\in N$.

There exists an elaborated theory of composition operators on various function spaces, presented in the monograph by Singh et al. \cite{singh1993composition}. In case $M, N$ are measure spaces, necessary as well as sufficient conditions on $\tau$  are given in \cite{singh1993composition}, Chapter $II$, and \cite{Takagi1999} under which $\tau$ generates a bounded linear operator $T_\tau : L^p(M)\rightarrow L^p(N) $. In the following we will only consider the case $p=2$.

We will discuss a simple example in order to illustrate the type of argument that arises in proving that a certain surjection $\tau$ generates a bounded linear operator $T_\tau$. 

\begin{rem}
In this example as well as also in the following parts of the paper, our notation will deviate from (\ref{def:T1}) and (\ref{def:T2}): In contrast to (\ref{def:T2}) we will define the bijective mapping as $$\tau : M\rightarrow N,$$ and the composition operator $T_\tau :\ L^2(M)\rightarrow L^2(N)$ by $T_\tau (f):= f\circ \tau^{-1}$. This simplifies some of the formulas and proofs, and it is in line with most papers on functional mappings.
\end{rem}

\begin{example}
Let $M=N$ denote the closed interval $[0,1]\in \mathbb{R}$ and let $\tau$ be a strictly monotonous differentiable real function mapping the interval onto itself with $t=\tau (s)$,  $0<\alpha \le\tau'(s)\le \beta$ on $[0,1]$. The condition on $\tau$  guarantees that $T_\tau$ and $T_{\tau}^{-1}$ exist and are bounded as  mappings from $L^2[0,1]$ into itself:

\begin{eqnarray*}
||T_\tau f||^2&=&\int_0^1f(\tau^{-1}(t))^2dt\\
&=&\int_0^1f(s)^2 \tau'(s)ds\\
&\le& \beta ||f||^2.
\end{eqnarray*}
The inverse composition operator $T_\tau^{-1}$ is also bounded: For $g\in L^2(0,1)$, we have $(T^{-1}_\tau g)(s) = g(\tau (s))$, thus
\begin{eqnarray*}
||T_\tau^{-1} g||^2&=&\int_0^1g(\tau(s))^2ds\\
&=&\int_0^1g(t)^2 \frac{dt}{\tau'(\tau^{-1}(t))} \\
&\le& \frac{1}{\alpha} ||g||^2.
\end{eqnarray*}

\end{example}

We now consider  two parametrized \emph{surfaces}  $M$, $N$ in $\mathbb{R}^3$. Let us assume that we are given differentiable global parametrizations  $s:U\rightarrow M$ and $t:V\rightarrow N$, where $s$ and $t$ are bijections of the open sets $U,V\in \mathbb{R}^2$.

 Let $L^2(M)$ and $L^2(N)$ denote the function spaces of square integrable function on $M$, $N$, respectively, which are separable \emph{Hilbert} spaces, equipped with the usual inner products and  norms based on the scalar surface integrals $\int_{M} fd\sigma_{M}$ and  $\int_{N} fd\sigma_{N}$ .

\begin{thm*}
Assume that $\tau: N \rightarrow M$ is a differentiable bijective map such that $$0<\alpha\le ||\tau'(s)|| \le \beta$$ for all $s\in M$.  Then $T_{\tau}: L^2(M)\rightarrow L^2(N)$ is a bounded linear operator with bounded inverse, and $$ ||T_\tau||\le \beta^2, ||T^{-1}_\tau||\le 1/\alpha^2 .$$
\end{thm*}

\begin{pro*}
Let us  define $g:= T_\tau ( f)$, which means $g(t)=f(\tau (t))$ for $t\in N$. Because of the bijectivity of $\tau$, the map $\sigma:=\tau\circ s:U\rightarrow M$ is also a differentiable parametrization of $N$. As the value of the surface integral is independent of the parametrization, this implies 
\begin{eqnarray*}
||T_\tau f||^2 &=& ||g||^2\\
&=&\int_N g^2 d \mu_N\\
&=&\int_U g(\tau (s(u)))^2 \sqrt{|det(G(u)^TG(u))|} du_1du_2\\
&=&\int_U f(s(u))^2 \sqrt{|det(A(u)^TB(u)^TB(u)A(u))|} du_1du_2
\end{eqnarray*}
where 
\begin{eqnarray*}
G(u)&=&\left(\frac{\partial \sigma_i}{\partial u_j}\right)_{\underset{j=1,2}{i=1,2,3}}\\
&=&\left(\frac{\partial \tau_i}{\partial s_k}\right)_{\underset{j=1,2,3}{i=1,2,3}} \left(\frac{\partial s_k}{\partial u_j}\right)_{\underset{j=1,2}{i=1,2,3}}\\
&=&B(u) A(u).
\end{eqnarray*}
Here $B=B(u)$ and $A=A(u)$ are $3 \times 3$ and $3 \times 2$ matrices, respectively. According to Theorem \ref{thm:Sperner} in Appendix \ref{app:1} (for $m$=3, $n$=2), 
\begin{eqnarray*}
||T_\tau f||^2&=&\int_U f((s(u)))^2 \sqrt{|det(A(u)^TB(u)^TB(u)A(u))|} du_1du_2\\
&\le &max_{u\in U}{||B(u)||^2} \int_U f((s(u)))^2 \sqrt{|det(A(u)^TA(u))|} du_1du_2\\
&=&max_{s\in M}{||\tau'(s)||^2} \int_M f^2 d \mu_M\\
&=& max_{s\in M}{||\tau'(s)||^2} \cdot ||f||^2.
\end{eqnarray*}
This shows that $T_\tau$ is bounded: $||T_\tau||\le  max_{s\in M}{||\tau'(s)||^2} \le \beta^2$. For the inverse $T^{-1}_\tau$, the same argument gives us $||T^{-1}_\tau||\le 1/(\min_{s\in M}||\tau'(s)||^2) \le 1/\alpha^2 $.

\begin{flushright}
$\square$
\par\end{flushright}

\end {pro*}

\begin{rem}
Singh and Manhas \cite{singh1993composition} treat composition operators in a very general setting:
Let  $(X, \mathcal{S}, m)$ be a measure space where $\mathcal{S}$ is a sigma algebra of subsets of $X$ and $m$ is a sigma-finite measure on $\mathcal{S},$ and let $L^p(X, \mathcal{S}, m)$ denote the set of all complex valued measurable functions $f$ on $X$ such that $|f|^p$ is integrable with respect to $m.$ The authors present conditions on the measure space and on $p\ge 1$ under which composition operators on $L^p(X,\Sigma,m)$ are well defined and bounded. The Radon-Nikodym derivative of measures plays an important role, generalizing the approach which we performed above in the one- and two-dimensional examples. The results of this important book are not directly applicable to problems with different shapes, as they only deal with composition operators from one $L^p$ space into itself. Generalizations to maps between different Lebesgue spaces $L^p(X, \mathcal{S}, m)$ and $L^q(X', \mathcal{S'}, m')$ have been given in \emph{Takagi et al.,} \cite{Takagi1999}.

\end{rem}

%%%%%%%%%%%%%%%%%%%%%%%%%%%%%%%%%%%%%%%%%%%%%%%%%%%%%

\section{Matrix representation of operators }
%%%%%%%%%%%%%%%%%%%%%%%%%%%%%%%%%%%%%%%%%%%%%%%%%%%%%

We consider two real separable Hilbert spaces $X$ and $Y$, and an operator $T: X\rightarrow Y$. In this section, we do not suppose that $T$ is a composition operator of type $T_\tau$ but just assume that $T$ is bounded with a bounded inverse. The idea of \emph{matrix representation} of linear operators is very old, see the textbook of  N. I. Akhiezer and I. M. Glazman, \cite{akhiezer2013theory}, p. 49, and it goes as follows. 

Let $\{\phi_k\}_{k\geq 1}$ and $\{\psi_k\}_{k\geq 1}$ be two orthonormal bases in $X$ and $Y$, respectively. Given $x\in X$, $y\in Y$ such that 
\begin{eqnarray}\label{eq:Tx=y}
Tx&=&y.
\end{eqnarray} 
Then
$$x=\sum_{k=1}^\infty x_k \phi_k,\hspace{1cm} y=\sum_{k=1}^\infty y_k \psi_k$$ with
$$ \sum_{k=1}^\infty |x_k|^2 < \infty,\hspace{1cm}  \sum_{k=1}^\infty |y_k|^2 < \infty.$$
Now the equation $Tx=y$ may be written as
\newline
\begin{center}
\begin{eqnarray*}
Tx&=&\sum_{i\ge 1} x_i T\phi_k\\
&=&\sum_{i\ge 1}x_i (\sum_{j\ge 1}(T\phi_i,\psi_j)\psi_j)\\
&=&\sum_{j\ge 1}(\sum_{i\ge 1}x_i (T\phi_i,\psi_j))\psi_j\\
&=&\sum_{j\ge 1}(\sum_{i\ge 1}x_i c_{ji})\psi_j\\
&=&\sum_{j\ge 1}y_j\psi_j\\
&=&y.
\end{eqnarray*}
\end{center}
Here we have made use of the continuity of the operator $T$, and we have defined 
$$c_{ij}:=(T\phi_j,\psi_i).$$ Thus we obtain the infinite linear system of equations
\begin{center}
\begin{eqnarray}\label{eq:cx=y}
 \begin{pmatrix}
c_{11} & c_{12} & c_{13} & \hdots\\
c_{21} & c_{22} & c_{23} & \hdots\\
c_{31} & c_{32} & c_{33} & \hdots \\
\hdots& \hdots& \hdots &\hdots
\end{pmatrix}
\begin{pmatrix}
x_1 \\
x_2 \\
x_3  \\
\vdots
\end{pmatrix}
&=&
\begin{pmatrix}
y_1 \\
y_2\\
y_3 \\
\vdots
\end{pmatrix}
\end{eqnarray}
\end{center}
This system represents a mapping between $\left\{x_k\right\}_{k=1}^{\infty}\in l_2$ and $\left\{y_k\right\}_{k=1}^{\infty}\in l_2$ ($l_2$ denotes the separable Hilbert space of all real sequences $\left\{a_k\right\}_{k=1}^{\infty}$ such that $\sum_{k=1}^\infty |a_k|^2 < \infty$). Because of $x_k=(x,\phi_k)$ and $y_k=(y,\psi_k),$  $(\ref{eq:cx=y})$ signifies that the doubly infinite matrix $C=(c_{ij})$ maps the sequence $\left\{x_k\right\}_{k=1}^{\infty}$ of generalized Fourier coefficients of $x$ to the generalized Fourier coefficients $\left\{y_k\right\}_{k=1}^{\infty}$ of $y$. 
The infinite matrix $C=(c_{ij})$ is called the \emph{matrix representation} of the linear operator $T$. In contrast to $T$ which maps a Hilbert space $X$ into a possibly different Hilbert space $Y$, the operator defined by $C$ maps $l_2$ into $l_2$ which simplifies the analysis. Abbreviating (\ref{eq:cx=y}), we also write 
\begin{equation}\label{eq:shortCx=y}
Cx=y,
\end{equation}
with a slight abuse of notation, as $x$ denotes an element of the separable Hilbert space $X$ as well as the corresponding element $(x_1,x_2,x_3....)\in l_2$, where $x_i=(x,\phi_i), i=1,2,...$. 

Not every infinite linear system (\ref{eq:cx=y}) represents a bounded operator on $l_2$, but there is a simple sufficient condition, the \emph{Schur test} (see \emph{Halmos} \cite{Halmos1982}, problem 45).

\begin{lem}{A sufficient condition:}
If  there are constants $\alpha\ge 0,\beta\ge 0$ such that  
$$\sum_{i=1}^{\infty} |c_{ij}| \le \alpha, j=1,2, ...$$
$$\sum_{j=1}^{\infty} |c_{ij}| \le \beta, i=1,2, ...$$
then(\ref{eq:cx=y}) defines a bounded operator on $l_2$ with $\|C\|\le \sqrt{\alpha \beta}.$
\end{lem}

There are refinements of the Schur test which make use of suitably defined weight functions for modifying the $L^p$-norms.This makes it possible to examine special classes of structured infinite matrices like band matrices etc. (see \cite{Lindner2008} and the literature cited there).

There is also a simple necessary condition which we will make use of later.

\begin{lem}{A necessary condition:}\label{lem:nesscond}
If  (\ref{eq:cx=y}) defines an operator on all of $l_2$, then there are constants $\gamma_j, j=1,2,...$ such that  
\begin{equation}\label{equ:nesscond}
\sum_{i=1}^\infty |c_{ij}|^2 \le \gamma_j , j=1,2, ...
\end{equation}
\end{lem}

\begin{pro*}
For $i=1,2,...$, let $e_i\in l_2$ be defined by $e_{ij}=\delta_{ij}$.  Then $Ce_i$ is the $i-th$ column of $C$, and because it is assumed to lie in $l_2$, (\ref{equ:nesscond}) holds true.

\begin{flushright}
$\square$
\par\end{flushright}
\end{pro*}

%%%%%%%%%%%%%%%%%%%%%%%%%%%%%%%%%%%%%%%%%%%%%%%%%%%
\section{The finite section method}
%%%%%%%%%%%%%%%%%%%%%%%%%%%%%%%%%%%%%%%%%%%%%%%%%%%

For this section we assume that the operator defined by the infinite matrix $C$ in (\ref{eq:cx=y}) is bounded as a map from $l_2$ to $l_2$, and that it has a bounded inverse, too.

\begin{nota*}
We regard elements of the Hilbert space $l_2$ as infinite column vectors $u=(u_1,u_2,\hdots)^T$, and we
%we make use of the identification
write
$\left( 
\begin{array}{c}
u\\
0
\end{array}
\right)
:=
(u^T,0,\hdots )^{T}\in l_2$ for any $u\in\mathbb{R}^n.$ The norm in $l_2$  is denoted by $\| \cdot\|_{l_2},$ and the norm in n-dimensional space $\mathbb{R}^n$ by $\|\cdot\|_n$. If $A$ is a linear map from $\mathbb{R}^n$ to $\mathbb{R}^m,$  $\|A\|_{n,m}$ denotes the operator norm. $\|\cdot \|_{\mathbb{R}^n\to \l_2} $ denotes the operator norm for a bounded linear operator which maps $\mathbb{R}^n $ into $l_2$.
\end{nota*}

We  are concerned with finite-dimensional approximations of the infinite linear system (\ref{eq:shortCx=y}). The most obvious approximation method is the following: For some fixed $n\ge 1$, consider the left upper $n\times n-$ submatrix $C^{(n,n)}$ of the infinite matrix $C$ and try to solve the linear system $$C^{(n,n)}x^{(n,n)}=y^{(n)}$$ which has as many variables as equations. Here $y^{(n)} \in\mathbb{R}^n$ is the first $n-$ section of $y\in l_2$, and $x^{(n,n)}\in \mathbb{R}^n$ is the approximation sought. This is called the (classical) \emph{Finite Section Method} which has been applied to many different types of equations, see \emph{Gr\"{o}chinig }et al. \cite{Groechenig2010}. But, as has been  shown by counterexamples (\cite{Boettcher1995}, \cite{Lindner2008}), this simple procedure is not always successful which means that the finite problems  do not always have solutions which approximate the solution of the infinite system. 
%%%%%%%%%%%%%%%%%%%%%%%%%%%%%%%%%%%%%%%%%%%
\subsection{Rectangular Finite Sections: The overdetermined case}
%%%%%%%%%%%%%%%%%%%%%%%%%%%%%%%%%%%%%%%%%%%
For some classes of infinite systems, there are different types of remedies to this situation. One of these is presented by Lindner ( see   Remark 3.5 of \emph{Lindner} \cite{Lindner2008} (see also \cite{lindner2006infinite}, \cite{LindnerHeinemeyer2008}) which amounts to solving, for $m\ge n$, a  least-squares problem with the \emph{rectangular} upper-left submatrix $C^{(m,n)}:$

\begin{equation}\label{prob:least-squares}
\min_{u\in \mathbb{R}^n  }\| C^{(m,n)}u-y^{(m)} \|_m.
\end{equation}

In case $m>n$, this finite problem amounts to solving an \emph{overdetermined} system of linear algebraic equations. For this task, there exist different very effective numerical methods (see Bj{\"o}rck \cite{Bjoerck1996}). - The theoretical investigation into the properties of this method (see \cite{Groechenig2006}, \cite{Lindner2008}) relies on the relation of the minimization problem (\ref{prob:least-squares}) to a least-squares problem in $l_2$:

Let $C^{(n)}$ be the semi-infinite matrix the columns of which are the first $n$ columns of the infinite matrix $C$, and let $x^{(n)}\in \mathbb{R}^n$ denote a solution of the linear least-squares problem
\begin{equation}\label{prob:least-squares_n}
\min_{v\in \mathbb{R}^n  }\| C^{(n)}v-y \|_{l_2}.
\end{equation}

\begin{thm}
 Let $C=(c_{ij})_{i,j\in\mathbb{N}}$ be the matrix representation of a bounded linear operator on $l_2$ with bounded inverse.
 \begin{enumerate}
 \item{For any sequence $\{x^{(n)}\}_{n\in\mathbb{N}}$ of solutions of  problem (\ref{prob:least-squares_n}),
$\lim_{n \to \infty} x^{(n)} = x,$
where $x$ is the unique solution of (\ref{eq:shortCx=y}).\footnote{
 According to our notational convention, $\lim_{n \to \infty} x^{(n)} = x$ means 
$
\left\| 
\left(
\begin{array}{c}
x{(n)}\\
0
\end{array}
\right)
-x
\right\|_{l_2}
\to 0
$ for 
$m\to\infty .$
}}
 \item{For any fixed $n\ge 1$, there is a $m_0(n)\in \mathbb{N}$ such that for all $m\ge m_0(n)$ the problem (\ref{prob:least-squares}) has a unique solution $x^{(m,n)} $. }
 \item{ For each $n\in\mathbb{N},$  $\lim_{m\to \infty} x^{(m,n)}=x^{n}$. }
 \item{There is a sequence $\{m_n\}_{n\in \mathbb{N}}$ such that $lim_{n\to \infty} x^{(m_n,n)}=x.$}
 \end{enumerate}
 
 \end{thm}
 
The idea of a proof is given in \cite{Lindner2008}, Remark 3.5 with reference to \cite{Groechenig2006}. For sake of completeness, we will here present a  proof of the theorem on different lines. We are going to show only the main road to the result, referring for technical details to Appendix 2.
\begin{pro*}
The proof of Assertion 1 is given in Lemma \ref{lem:1}.
In order to show Assertion 2, we will prove that, for every  $n\ge 1$, there is a $m_0(n)\ge 1$ such that $C^{(m,n)}$ has full rank $n$ for all $m\ge m_0(n)$.
Assume that there is a  $n\ge 1$ such that there exists a sequence $\{m_k\}_{k\in  \mathbb{N}}$ and 
$\{u_k\}_{k\in \mathbb{N}}\in \mathbb{R}^n$ such that $\left\|u_k\right\|_n=1$ und $C^{(m_k,n)}u_k=0$. $\{u_k\}_{k\in \mathbb{N}}$
has a convergent subsequence $u_{k_j}\to u$ for $j\to \infty$, satifying $\left\|u\right\|_n=1$ . This implies $C^{(n)} u=0$ and thus $C \left(
\begin{array}{c}
 u \\
 0 \\
\end{array}
\right)=0$, which contradicts the bijectivity of $C$. Thus, Assertion 2 holds true. In order to prove Assertion 3,
 we will show first that, for every $n\ge 1$, $\|x^{(m,n)}\|_n$ is bounded independent of $m$.
 As $x^{(m,n) }$ is a solution of (\ref{prob:least-squares}),

\begin{eqnarray*}
\left\|C^{(m,n)}x^{(m,n)}-y^{(m)}\right\|_m&\leq& \left\|C^{(m,n)}x^{(n)}-y^{(m)}\right\| _m\\
&\leq &\left\|C^{(m,n)}\right\|_{n,m} \|x^{(n)}\|_n+\left\|y^{(m)}\right\| _m\\
&\leq&\left\|C^{(n)}\right\|_{\mathbb{R}^n\to \l_2} \|x^{(n)}\|_n+\|y\|_m
\end{eqnarray*}

and therefore, by Lemma \ref{lem:2},

\begin{eqnarray*}
\mu (n)\left\|x^{(m,n)}\right\|_n&\leq& \left\| C^{(m,n)}x^{(m,n)}\right\|_m\\
& \leq&  \left\|C^{(m,n)}x^{(m,n)}-y^{(m)}\right\|_m+\left\|y^{(m)}\right\|_m \\
&\leq&\left\|C^{(n)}\right\|_{\mathbb{R}^n\to \l_2} \|x^{(n)}\|_n+2\|y\|_{l_2}
\end{eqnarray*}

for all $m\geq m_0(n)$, which implies the boundedness of $\left\|x^{(m,n)}\right\|$:

\begin{eqnarray*}
\left\|x^{(m,n)}\right\|_n&\leq &\frac{1}{\mu^{(n)}}\left(\left\|C^{(n)}\right\|_{\mathbb{R}^n\to \l_2} \|x^{(n)}\|_n+2\|y\|_m\right), 
\end{eqnarray*}

Let $K^{(m,n)}$ and $z^{(m)}$ be defined by
$C^{(n)}=\left(
\begin{array}{c}
 C^{(m,n)} \\
 K^{(m,n)} \\
\end{array}
\right)$ und $y=\left(
\begin{array}{c}
 y^{(m)} \\
 z^{(m)} \\
\end{array}
\right).$ 
where $y^{(m)}\in \mathbb{R}^n$ and $z^{m}\in l_2$.
Applying Lemma \ref{lem:opt} of Appendix \ref{app:over} to  Problem (\ref{prob:least-squares_n}), we obtain

\begin{eqnarray}
\left\|C^{(n)}\left(x^{(m,n)}-x^{(n)}\right)\right\|_{l_2}^2&=&\left\|C^{(n)}x^{(m,n)}-y\right\|_{l_2}^2-\left\|C^{(n)}x^{(n)}-y\right\|_{l_2}^2\\
&=&\left\|C^{(m,n)}x^{(m,n)}-y^{(m)}\right\|_m^2-\left\|C^{(m,n)}x^{(n)}-y^{(m)}\right\|_{l_2}^2\\
&+&\left\|K^{(m,n)}x^{(m,n)}-z^{(m)}\right\|_{l_2}^2-\left\|K^{(m,n)}x^{(n)}-z^{(m)}\right\|_{l_2}^2\\
&\le&\left\|K^{(m,n)}x^{(m,n)}-z^{(m)}\right\|_{l_2}^2-\left\|K^{(m,n)}x^{(n)}-z^{(m)}\right\|_{l_2}^2
\end{eqnarray}
The last inequality holds true because $x^{(m,n)}$ is a minimizer of  problem (\ref{prob:least-squares_n}).
%&\leq &\right\|K^{(m,n)}x^{(m,n)}-z^{(m)}\|^2-\left\|K^{(m,n)}x^{(n)}-z^{(m)}\|^2\right.\right.\right.\right.
 Thus we have proved $\left\|C^{(n)}\left(x^{(m,n)}-x^{(n)}\right)\right\|_{l_2}\to 0 $ for $ m\to \infty$ because, by Lemma \ref{lem:nesscond}, $\left\|K^{(m,n)}\right\|_{\mathbb{R}^n\to \l_2}\to 0$, because of the boundedness of $ \left\|x^{(m,n)}\right\|_n$ and because of $\left\|z^{(m)}\right\|_{l_2}\to 0 $ for $ m\to \infty   $. Therefore

$$\left\|x^{(m,n)}-x^{(n)}\right\|_n
\leq \frac{1}{\mu ^{(n)}}\|C^{(n)}\left(x^{(m,n)}-x^{(n)}\right)\|_{l_2}\to 0, m\to \infty .$$ Assertion 4 follows directly from 1 and 3.

\begin{flushright}
$\square$
\par\end{flushright}

\end{pro*}

%%%%%%%%%%%%%%%%%%%%%%%%%%%%%%%%%%%%%%%%%%%%%
\subsection{Rectangular Finite Sections: The underdetermined case}
%%%%%%%%%%%%%%%%%%%%%%%%%%%%%%%%%%%%%%%%%%%%%
Instead of using approximations by overdetermined linear systems, we will propose a new method which leads to the computation of minimum-norm solutions of \emph{underdetermined} systems: Instead of fixing the first $n$ \emph{columns} of $C$ and   "cutting" them after row $m$ where $m>n$, one can fix the first $m$ \emph{rows} of $C$ and cut them after column $n$, where (where $m\le n$). This will produce an underdetermined linear system $$C_{(m,n)}x_{(m,n)}=y_{(m)}$$ 
For $m,n\in \mathbb{N}$,  $m\leq n$, $C_{(m,n)}$ denotes the $m\times n$  upper left submatrix .   $y_{(m)}\in \mathbb{R}^m$ is the vector consisting of the first  $m$ components of $y$. 

We propose to compute $x_{(m,n)}$ as the unique \emph{minimum-norm solution} of this linear system, i.e., as the solution of the quadratic optimization problem

\begin{equation}\label{prob:underdeter}
 \min_{u\in \mathbb{R}^n }  \left\{ \|u\|_n^2 \:|\: C_{(m,n)}u=y_{(m)} \right\}
\end{equation}

In analogy to the overdetermined case, we introduce an "intermediate" minimum-norm problem: For each  $m\in \mathbb{N}$ let $C_{(m)}$ denote the semi-infinite submatrix of $C$ consisting of the first $m$ rows of $C$.  $x_{(m)}\in l_2$ denotes the unique solution of the minimum-norm problem

\begin {equation}\label{prob:C(m)}
\min_{u\in l2} \{\| u \|_{l_2}^2 \: | \:    C_{(m)}u=y_{(m)}\}.
\end{equation}

\begin{rem*}
Of course $C^{(m,n)}=C_{(m,n)} $ for all $m,n\in \mathbb{N}$, and $y^{{m}}=y_{(m)}$ for all $m$. For sake of clarity we use different notation in the overdetermined case and in the underdetermined one, respectively. Of course  $C^{(m)}$ and $C_{(n)} $ are different for all $ m,n\in \mathbb{N}$:  $C^{(n)}$ has an infinite number of rows (and a number of  $n$ columns), while $C_{(m)}$ has an infinite number of columns (and $m$ rows).
\end{rem*}   

\begin{thm}
 Let $C=(c_{ij})_{i,j\in\mathbb{N}}$ be the matrix representation of a bounded linear operator on $l_2$ with bounded inverse.
 \begin{enumerate}
 \item{ $\lim_{m \to \infty} x_{(m)} = x,$ where $x$ is the unique solution of (\ref{eq:shortCx=y}}) and $\{x_{(m)}\}_{m\in\mathbb{N}}$ is the sequence of solutions of  problems (\ref{prob:C(m)}).
 \item{For any fixed $m\ge 1$,  the problems (\ref{prob:underdeter}) have unique solutions $x_{(m,n)} $ for all $n\ge m.$ }
 \item{ For each $m\in\mathbb{N},$  $\lim_{n\to \infty} x_{(m,n)}=x_{m}$. }
 \item{There is a sequence $\{n_m\}_{m\in \mathbb{N}}$ such that $lim_{m\to \infty} x_{(m,n_m)}=x.$}
 \end{enumerate}
 \end{thm}
 
 \begin{pro*}
  Again we present the main line of the proof, shifting some technical lemmas to Appendix \ref{app:under}.- Assertion 1: By Lemma \ref{lem:opt2}, 
$$\|(x-x_{(m)})\|_{l_2}^2=\|x\|_{l_2}^2-\|x_{(m)}\|_{l_2}^2\ ,$$
so we have to show only $\|x_{(m)}\|_{l_2}\to \|x\|_{l_2}$. 
$x$ is admissible for all optimization problems (\ref{prob:C(m)}), and each $x_{(m)}$ is admissible for all "preceding" problems with smaller index $m$ , thus
$$\left\|x_{(1)}\right\|_{l_2}\leq \left\|x_{(2)}\right\|_{l_2}\leq \left\|x_{(3)}\right\|_{l_2}\leq  \ldots \leq \|x\|_{l_2},$$
therefore
$\lim_{m\to \infty } \left\|x_{(m)}\right\|_{l_2}=:r\leq \|x\|_{l_2}$. \emph{Assumption:} $r<\|x\|_{l_2}$. Then   $x_{(m)}$ has a subsequence $x_{(m_k)}$ converging weakly to $\tilde{x}$ with $\left\|\tilde{x}\right\|_{l_2}\leq r$. This implies, for alle $u\in l_2$,
\begin{eqnarray*}
\left(Cx_{(m_k)},u\right)&=&\left(x_{(m_k)},C^*u\right)\\
&\to& \left(\tilde{x},C^*u\right)\\
&=&\left(C\tilde{x},u\right), k\to \infty.
\end{eqnarray*}
We now choose  $u$ as unit vector $u_i$, which implies
$$\left(Cx_{(m_k)}\right)_i\to \left(C\tilde{x}\right)_i$$ for all $i\in \mathbb{N}$, $k\to \infty$.
But as $\left(Cx_{(m_k)}\right)_i=y_i$ as soon as $m(k)\geq i$, it follows that $\left(C\tilde{x}\right)_i=y_i$ for all $i\in \mathbb{N}$, implying
$C \tilde{x}=y.$  Therefore it follows that $\tilde{x}=x$, in contrast to the assumption  $\left\|\tilde{x}\right\|_{l_2}\leq r<||x\|_{l_2}$. Thus $\left\|x_{(m)}\right\|_{l_2}\to \|x\|_{l_2}$, implying $\left\|x-x_{(m)}\right\|_{l_2}\to 0$ for $m\to \infty$, and the first assertion of the theorem is proved.

Assertion 2: The uniqueness of $x_{(m,n)}$ follows from the fact that problems (\ref{prob:least-squares}) are convex optimization problems having a uniformly convex cost function. 

Assertion 3: Let $m$ be fixed. The boundedness of $\left\|\left(C_{(m,n_k)}C_{(m,n_k)}^*\right)^{-1}\right\|_{m,m}$ (see Lemma \ref{lem:last}) and $z_{(m,n)}=\left(C_{(m,n)}C_{(m,n)}^*\right){}^{-1}y^{(m)}$ implies that the sequence $z_{(m,n)}$ is bounded. Let $z_{(m,n_k)}\to\tilde{z}$ be a convergent subsequence, for which it follows that
\begin{eqnarray*}
C_{(m)}C_{(m)}^*z_{(m,n_k)}&=&\left(C_{(m)}C_{(m)}^*-C_{(m,n_k)}C_{(m,n_k)}^*\right)z_{(m,n_k)}\\
&+&C_{(m,n_k)}C_{(m,n_k)}^*z_{(m,n_k)}\\
&=&\left(C_{(m)}C_{(m)}^*-C_{(m,n_k)}C_{(m,n_k)}^*\right)z_{(m,n_k)}+y_{(m)}
\end{eqnarray*}
It follows from Assertion 4 of Lemma \ref{lem:last}, that $C_{(m)}C_{(m)}^*\tilde{z}=y^{(m)}$ and therefore $\tilde{z}=z_{(m)}$. As each convergent subsequence of $z_{(m,n)}$ converges to $z_{(m)}$ it follows that the whole sequence $z_{(m,n)}$ converges to $z_{(m)}$.

\begin{flushright}
$\square$
\par\end{flushright}

 \end{pro*}

The solution of the minimum-norm problem (\ref{prob:underdeter}) can be written as 
$$x_{(m,n)}=C_{(m,n)}^+y_{(m)},$$
and there exist efficient numerical algorithms for computing this solution.

\section{Summary}
In this paper we are concerned with the  foundational aspects of the method of "Functional maps" for describing and computing maps between shapes. Our aim is to draw attention to the mathematical structure of the underlying well established theories of composition maps between Hilbert spaces, matrix representation of operators, and the finite section method. Concerning the latter, we  pointed to the fact that     the standard "quadratic" finite section  method cannot guarantee convergence. In addition to the known "rectangular" modification which leads to systems of overdetermined linear algebraic equations, we describe a new variant which produces underdetermined systems of which the minimum norm solution has to be computed. - We hope that these theoretical considerations stimulate the activity in this new field of application of mathematics in computer graphics and that this may lead to new insights into existing algorithms  or even design of new ones.

%\input{sections/appendix.tex}

%%%%%%%%%%%%%%%%%%%%%%%%%%%%%%%%%%%%%%%%%%%%%
\section{Appendix: Proof of theorems}
%%%%%%%%%%%%%%%%%%%%%%%%%%%%%%%%%%%%%%%%%%%%%

\subsection{Appendix 1: A theorem on determinants}\label{app:1}
%%%%%%%%%%%%%%%%%%%%%%%%%%%%%%%%%%%%%%%%%%%%%
\begin{thm}\label{thm:Sperner}
 Let $A,B$ be real $m\times n$ and $n\times n-$ matrices, respectively, and let  $d_1\geq d_2\geq \ldots \geq d_m$ be the singular values of $B$. Then it holds that
\begin{equation}\label{ineq12}
 \det \left(A^TB^TB A\right)\leq d_1{}^2*\ldots *d_n{}^2*\det \left(A^TA\right)\leq \|B\|_2^{2n}*\det \left(A^TA\right)
\end{equation}
\end{thm}

\begin{pro*}

Let the singular value decomposition of $B$ be  given: $B=U D V^T$ where $U, V$ are $n\times n$ orthogonal, and $D=diag(d_1,d_2,...,d_n)$ is diagonal.
Defining $C:=V^TA$, we obtain
$\det \left(A^TB^TB A\right)=\det \left(A^TV D^2V^TA\right)=\det \left(C^TD^2C\right)$ 
und $\det \left(A^TA\right)=\det \left(A^TV V^TA\right)=\det \left(C^TC\right)$.
So all we have to show in order to prove the first inequality is
\begin{equation*}
\det \left(C^TD^2C\right)\leq d_1{}^2*\ldots *d_n{}^2*\det \left(C^TC\right).
\end{equation*}
Let $C=\left(c_{i j}\right), i=1,...,m; j=1,...,n.$  
According to a classical theorem of the theory of determinants (\emph{E. Sperner} \cite{Sperner1965}, p.194; see also Frohman \cite{Frohman2010} \footnote{In his Epilogue, Frohman writes: "This is a theorem that gets rediscovered over and over again."}, who called it  "Full Theorem of Pythagoras"),
\begin{equation*}
\det \left(C^TC\right)=\sum _{1\leq i_1<\ldots <i_n\leq m} \left(\det \left(
\begin{array}{ccc}
 c_{i_11} & \ldots  & c_{i_1n} \\
 \vdots  &   & \vdots  \\
 c_{i_n1} & \ldots  & c_{i_nn} \\
\end{array}
\right)\right)^2
\end{equation*}
and accordingly

\begin{eqnarray*}
\det \left(C^TD^2C\right)&=&\det \left((D C)^T(D C)\right)\\
&=&\sum _{1\leq i_1<\ldots <i_n\leq m} \left(\det \left(
\begin{array}{ccc}
 d_{i_1}c_{i_11} & \ldots  & d_{i_1}c_{i_1n} \\
 \vdots  &   & \vdots  \\
 d_{i_n}c_{i_n1} & \ldots  & d_{i_n}c_{i_nn} \\
\end{array}
\right)\right)^2\\
&=&\sum _{1\leq i_1<\ldots <i_n\leq m} \left(d_{i_1}*\ldots *d_{i_n*}\det \left(
\begin{array}{ccc}
 c_{i_11} & \ldots  & c_{i_1n} \\
 \vdots  &   & \vdots  \\
 c_{i_n1} & \ldots  & c_{i_nn} \\
\end{array}
\right)\right)^2\\
&\leq&  d_1{}^2*\ldots *d_n^2*\sum _{1\leq i_1<\ldots <i_n\leq m} \left(\det \left(
\begin{array}{ccc}
 c_{i_11} & \ldots  & c_{i_1n} \\
 \vdots  &   & \vdots  \\
 c_{i_n1} & \ldots  & c_{i_nn} \\
\end{array}
\right)\right)^2\\
&=&d_1{}^2*\ldots *d_n{}^2*\det \left(C^TC\right)
\end{eqnarray*}
which proves the first inequality of (\ref{ineq12}). The second inequality holds because of $\|B\|_2=d_1\ge d_2 \ge ... \ge d_n$.

\begin{flushright}
$\square$
\par\end{flushright}

\end{pro*}

\subsection{Appendix 2: Rectangular Finite Sections (the overdetermined case}\label{app:over}
%%%%%%%%%%%%%%%%%%%%%%%%%%%%%%%%%%%%%%%%%%%%
\begin{lem}\label{lem:1}
For any sequence $\{x^{(n)}\}_{n\ge 1}$ of solutions of  problem (\ref{prob:least-squares_n}),
$\lim_{n \to \infty} x^{(n)} = x,$
where $x$ is the unique solution of (\ref{eq:shortCx=y}).
\end {lem}

\begin{pro*}
\begin{eqnarray}
\left\|C \left(\left(
\begin{array}{c}
 x^{(n)} \\
 0
\end{array}
\right)-x\right)\right\|_{l_2}&=&\left\|C^{(n)}x^{(n)}-y\right\|_{l_2} \\
&\leq& \left\|C^{(n)}\left(
\begin{array}{c}
 x_1 \\
 \vdots  \\
 x_n \\
\end{array}
\right)-y\right\|_{l_2} \\
&=&\left\|C \left(\left(
\begin{array}{c}
 x_1 \\
 \vdots  \\
 x_n \\
 0 
\end{array}
\right)-x\right)\right\|_{l_2}\to 0 
\end{eqnarray}
for $n\to \infty$

As $C$ is assumed to have a bounded inverse,

$\left\|\left(
\begin{array}{c}
 x^{(n)} \\
 0  
\end{array}
\right)-x\right\|_{l_2}
=
\left\|
C^{-1}(C\left(
\begin{array}{c}
 x^{(n)} \\
 0 
\end{array}
\right)-y)\right\|_{l_2} 
\to 0 $  for $n\to \infty$.

\begin{flushright}
$\square$
\par\end{flushright}

\end{pro*}

\begin{lem}\label{lem:2}
\begin{enumerate}
\item{$\left\| C^{(m,n)}\right\|_{n,m} \le   \left\| C^{(n)}\right\|_{\mathbb{R}^n\to \l_2}$ for all $m\ge 1$};
\item{For each $n\ge 1$ there is a $m_0(n)\ge 1$ and a $\mu^{(n)}$ such that $$\mu ^{(m,n)}\text{:=}\min \left\{\left\|C^{(m,n)}u\right\|_m:u\in \mathbb{R}^n,\|u\|_n=1\right\}>\mu^{(n)}>0$$ for all $m\geq m_0(n)$.}
\end{enumerate}
\end{lem}

\begin{pro*}
\begin{enumerate}
\item{
Because $\left\|C^{(m,n)}u\right\|_m\leq \left\|C^{(m+1,n)}u\right\|_{m+1}\leq \left\|C^{(n)}u\right\|_{l_2}$ for all $u\in\mathbb{R}^n,$
\begin{eqnarray*}
\left\|C^{(m,n)}\right\|_{n,m}&=&\max \left\{\left\|C^{(m,n)}u\right\|_m:u\in \mathbb{R}^n,\|u\|_n=1\right\}\\
&\leq &\max \left\{\left\|C^{(n)}u\right\|_{l_2}:u\in \mathbb{R}^n,\|u\|_n=1\right\}\\
&=&\left\|C^{(n)}\right\|_{\mathbb{R}^n\to \l_2}.
\end{eqnarray*}
}

\item{We will prove first that, for every  $n\ge 1$, there is a $m_0(n)\ge 1$ such that $C^{(m,n)}$ has full rank $n$ for all $m\ge m_0(n)$.
Assume that there is a  $n\ge 1$ such that there exists a sequence $\{m_k\}_{k\in  \mathbb{N}}$ and 
$\{u_k\}_{k\in \mathbb{N}}\in \mathbb{R}^n$ such that $\left\|u_k\right\|_n=1$ und $C^{(m_k,n)}u_k=0$. $\{u_k\}_{k\in \mathbb{N}}$
has a convergent subsequence $u_{k_j}\to u, \left\|u\right\|_n=1,$ for $j\to \infty$. This implies $C^{(n)} u=0$ and thus $C \left(
\begin{array}{c}
 u \\
 0 
\end{array}
\right)=0$ which contradicts the bijectivity of $C$.
}
The full rank of $C^{(m,n)}$ for $m\ge m_0$ implies
$\mu ^{(m,n)}\text{:=}\min \left\{\left\|C^{(m,n)}u\right\|_m:u\in \mathbb{R}^n,\|u\|_n=1\right\}>0$ for all $m\geq m_0(n)$. As

$\mu ^{(m,n)}$ is monotone non decreasing in $m$,  $\left\|C^{(m,n)}u\right\|_m\geq \mu ^{(m_0,n)}\|u\|_n$ for all $u\in \mathbb{R}^n$ and $m\ge m_0(n)$. Therefore, \ref{lem:2} holds true with $\mu^{(n)}:=\mu^{(m_0,n)}$.

\end{enumerate}
\begin{flushright}
$\square$
\par\end{flushright}
\end{pro*}

The following Lemma is a well known result of the theory of least-squares problems. 

\begin{lem}\label{lem:opt}
Let there be given a real $m\times n-$ matrix $A$ $(m\ge n)$ and a vector $b\in\mathbb{R}^m$, and consider
 the least-squares problem
$$\|A w-b\|_m=\min !, w\in \mathbb{R}^n.$$  
$w_0\in \mathbb{R}^n$ is a solution of this problem if and only if, for any $w\in \mathbb{R}^n$,
$$\|A w-b\|_m^2=\|A\left(w-w_0\right)\|_m^2+\|A w_0-b\|_m^2.$$
\begin{pro*}
A necessary condition for $w_0$ to be the solution of the above problem is the system of normal equations $A^T\left(A w_0-b\right)=0.$ This implies, for all $w\in \mathbb{R}^n$, $\left(A w_0-b,A\left(w-w_0\right)\right)=\left(A^*\left(A w_0-b\right),w-w_0\right)=0$  and therefore
\begin{eqnarray*}
\|A w-b\|_m^2&=&\|A\left(w-w_0\right)+\left(A w_0-b\right)\|_m^2\\
&=&\|A\left(w-w_0\right)\|_m^2+\left(A w_0-b,A\left(w-w_0\right)\right)+ \|A w_0-b\|_m^2\\
&=&\|A\left(w-w_0\right)\|_m^2+\|A w_0-b\|_m^2 
\end{eqnarray*}
The sufficiency of the condition follows from  
$$\|A w_0-b\|_m^2  =\|A w-b\|_m^2 - \|A\left(w-w_0\right)\|_m^2\le\|A w-b\|_m^2 $$
for all $w\in \mathbb{R}^n.$
\begin{flushright}
$\square$
\par\end{flushright}
\end{pro*}

\end{lem}

\subsection{Appendix 3: Rectangular Finite Sections: The underdetermined case}\label{app:under}
%%%%%%%%%%%%%%%%%%%%%%%%%%%%%%%%%%%%%%%%%%%%%%%

\begin{lem}\label{lem:opt2}
There is a unique solution of $x_{(m)}$ of (\ref{prob:C(m)}) for any $m\in \mathbb{N}$, and it is characterized by the existence of a unique vector $z_{(m)}\in \mathbb{R}^n$ such that
$$x_{(m)}=C_{(m)}^*z_{(m)}$$ and $$C_{(m)}C_{(m)}^*z_{(m)}=y_{(m)}.$$
For any $u$ such that $C_{(m)}u=y_{(m)}$ it holds that
$$\|(u-x_{(m)})\|_n^2=\|u\|_n^2-\|x_{(m)}\|_n^2$$.
\end{lem}
\begin{pro*}
It is well known that $x_{(m)}$ is the unique solution of (\ref{prob:C(m)}), if and only if $x_{(m)}$ is orthogonal to the nullspace of $C_{(m)}$ which is the same as the range space of $C_{(m)}^*$. (The uniqueness of $x_{(m)}$ follows from the fact that it is the solution of a convex optimization problem having a uniformly convex cost function.) This explains the first equation of the Lemma. The second equation just says that $x_{(m)}$ is admissible to problem (\ref{prob:C(m)}), i.e. it  solves the underdetermined linear system $C_{(m)}u=y_{(m)}$. The uniqueness of $z_{(m)}$ follows because $C_{(m)}C_{(m)}^*$ has full rank $m$. The last equality follows from
\begin{eqnarray*}
\|u\|_n^2&=&\|(u-x_{(m)})+x_{(m)}\|_n^2\\
&=&\|(u-x_{(m)})\|_n^2+(u-x_{(m)},x_{(m)})+\|x_{(m)}\|_n^2\\
&=&\|(u-x_{(m)})\|_n^2+(C_{(m)}^*z_{(m)},x-x_{(m)})+\|x_{(m)}\|_n^2\\
&=&\|(u-x_{(m)})\|_n^2+(z_{(m)},C_{(m)}(x-x_{(m)}))+\|x_{(m)}\|_n^2\\
&=&\|(u-x_{(m)})\|_n^2+\|x_{(m)}\|_n^2
\end{eqnarray*}

\begin{flushright}
$\square$
\par\end{flushright}

\end{pro*}

\begin{lem}\label{lem:last}
Let $C_{(m,n)}^*$denote the transposed matrices of $C_{(m,n)}$, and let $C_{(m)}^*$ denote the adjoint of $C_{(m)}$. Then
\begin{enumerate}
\item{The operator norms of $C_{(m,n)}^*$ are bounded, and  $\left\|C_{(m,n)}^*\right\|_{n,m}\leq \left\|C_{(m)}^*\right\|_{l_2\to\mathbb{R}^m}$ for all $n\in \mathbb{N}.$}
\item{$\mu _{(m,n)}\text{:=}\min \left\{\left\|C_{(m,n)}^*w\right\|_n:w\in \mathbb{R}^m,\|w\|_m=1\right\}\ge \mu_{(m)}>0$ for all $n\ge n_0(m).$}
\item{$\left\|\left(C_{(m,n)}C_{(m,n)}^*\right){}^{-1}\right\|_{m,m}\leq 1\left/\mu _{(m,n)}^2\right.\leq 1\left/\mu _{\left(m,n_0\right)}^2\right.$ for all $n\geq n_0$}
\item{For fixed $m$, $$\|C_{(m)}C_{(m)}^*-C_{(m,n)}C_{(m,n)}^*\|_{m,m}\to 0$$ for $n\to \infty$.}
\end{enumerate}
\end{lem}

\begin{pro*}
Assertions 1 and 2: Exactly like in the proof of Lemma \ref{lem:2}, one can prove the corresponding properties for the transposed matrices $C_{(m,n)}^*$ and the adjoint $C_{(m)}^*$. Assertion 3 follows from Assertion 2.  Assertion 4 follows from the necessary condition of Lemma \ref{lem:nesscond}.

\begin{flushright}
$\square$
\par\end{flushright}

\end{pro*}

\bibliographystyle{plain}
\bibliography{biblio}

\end{document}